\documentclass[11pt,reqno]{amsart}
\usepackage[colorinlistoftodos,shadow]{todonotes}
\usepackage{amsmath,amscd,amssymb}
\usepackage[T1]{fontenc}
\usepackage[utf8]{inputenc}
\usepackage{cite}
\usepackage{color}
\usepackage{graphicx}
\usepackage{psfrag,epsfig}
\usepackage{epstopdf}
\usepackage{multirow}
\usepackage{graphicx}
\usepackage{rotating}
\usepackage{fullpage}
\usepackage[letterpaper,margin=1.0in]{geometry}
\usepackage{subfigure}
\usepackage{enumerate}
\usepackage{comment}
\usepackage{lineno}
\usepackage{algorithm2e}
\usepackage[small]{caption}
\usepackage{graphicx}
\usepackage[super]{natbib}
\usepackage{tabu} 
\usepackage{float}
\usepackage{mathrsfs}
\usepackage{amssymb}
\usepackage{amsmath}
\usepackage{mathtools}
\usepackage[debug=false, colorlinks=true, pdfstartview=FitV, 
linkcolor=blue, citecolor=blue, urlcolor=blue]{hyperref}

\newcommand{\vect}[1]{\mathbf{#1}}

\newcommand{\gradient}[1]{\ensuremath{\nabla{#1}}}
\newcommand{\divergence}[1]{\ensuremath{\nabla\cdot{#1}}}

\newcommand{\vx}[1]{\vect{x{#1}}}
\newcommand{\dt}[1]{\Delta t{#1}}

\numberwithin{equation}{section}
\graphicspath{{./}}
\newcommand*\patchAmsMathEnvironmentForLineno[1]{%
  \expandafter\let\csname old#1\expandafter\endcsname\csname #1\endcsname
  \expandafter\let\csname oldend#1\expandafter\endcsname\csname end#1\endcsname
  \renewenvironment{#1}%
     {\linenomath\csname old#1\endcsname}%
     {\csname oldend#1\endcsname\endlinenomath}}%
\newcommand*\patchBothAmsMathEnvironmentsForLineno[1]{%
  \patchAmsMathEnvironmentForLineno{#1}%
  \patchAmsMathEnvironmentForLineno{#1*}}%
\AtBeginDocument{%
\patchBothAmsMathEnvironmentsForLineno{equation}%
\patchBothAmsMathEnvironmentsForLineno{align}%
\patchBothAmsMathEnvironmentsForLineno{flalign}%
\patchBothAmsMathEnvironmentsForLineno{alignat}%
\patchBothAmsMathEnvironmentsForLineno{gather}%
\patchBothAmsMathEnvironmentsForLineno{multline}%
}

\title[Large-scale Inversion of Subsurface Flow]{Large-scale Inversion of Subsurface Flow using Discrete Adjoint Method}
\author[S.~Wang et al.]{S.~Wang$^{1,2}$, S.~Karra$^{3,*}$ and D.~O'Malley$^{3}$\\
{\scriptsize $^{1}$Department of Electrical Engineering, 
University of New Mexico, Albuquerque, NM 87131\\ }
{\scriptsize $^{2}$National Security Education Center, Los Alamos National 
Laboratory, Los Alamos, NM 87545.\\}
{\scriptsize $^{3}$Computational Earth Science Group,
Earth and Environmental Sciences Division, 
Los Alamos National Laboratory, Los Alamos, NM 87545.} }
\thanks{$^*$Corresponding author, \texttt{satkarra@lanl.gov}}
\date{\today}
\begin{document}
\maketitle
\tableofcontents
\clearpage
\nolinenumbers
%
\section*{Abstract} 

\textbf{Keywords:}~subsurface, flow, inversion, adjoint method, sensitivity analysis, parallel, high performance computing. 
%
\section{Introduction}

Inverse analysis plays a key role in developing realistic models for subsurface hydrogeology.
This is largely because state variables such as pressure can be observed, but constitutive parameters such as permeability that are needed to make predictions can only be inferred from observations of the state variables.
There is a long history of applying inverse methods in subsurface hydrology \citep{neuman1980statistical1, neuman1980statistical2, Carrera-1986-Estimation, Sun-1994-Inverse, Kitanidis-1997-Introduction, zhang1997iterative, carrera2005inverse} often via the geostatistical approach\citep{kitanidis-1995-Quasi, zhang1997iterative, Kitanidis-1997-Introduction, Kitanidis-1997-Minimum, Vesselinov2001i} and using a variety of computational techniques such as dimension reduction\cite{lee2014large}, subspace recycling\cite{lin2016computationally}, and even quantum computational methods\cite{o2018approach}.
Recently, it is increasingly important to calibrate the model to match large data sets obtained from observing pressure transients at a relatively modest number of wells over a long period of time\cite{lin2017large}.

As a result of an inverse analysis, subsurface hydrologic modelers can often use these calibrated models to make accurate predictions related to, e.g., the impacts of pumping at one well on the water supply at another well or the fate of contaminants in groundwater \cite{o2014combined}.
In other cases, the data that has been used to calibrate the model may not be sufficient to use the model in a predictive fashion.
When this happens, the calibrated model is often used as a starting point for an uncertainty analysis (e.g., as the starting point in a Markov Chain Monte Carlo or in a null space Monte Carlo method \cite{tonkin2009calibration}).
Often these analyses are used to inform decisions (e.g., related to remediating contaminated groundwater \cite{o2014groundwater}).
\section{Formulation}
Let $\Omega \subset \mathbb{R}^{\textit{nd}}$ be a bounded open domain, where ``\textit{nd}'' is the number of spatial dimensions. The boundary $\partial \Omega=\bar{\Omega}-\Omega$ is assumed to be piecewise smooth. The boundary is divided into two parts: $\Gamma^D$ and $\Gamma^N$. $\Gamma^D$ ($\Omega^N$)is that part of the boundary on which Dirichlet(Neumann) boundary conditions are prescribed. For mathematical well-posedness, we assume $\Gamma^D \cup \Gamma^N=\partial \Omega$ and $\Gamma^D \cap \Gamma^N=\emptyset$. The unit outward normal to boundary is denoted as $\hat{\vect{n}}$. The permeability tensor is denoted by $\vect{D}(\vx)$, which is assumed to symmetric, bounded above and uniformly elliptic. That is, there exists two constant $0<\epsilon_1 \leq \epsilon_2 < \infty$ such that
\begin{equation}
\epsilon_1 \vect{y}^\mathrm{T}\vect{y}\leq \vect{y}^\mathrm{T}\vect{D}(\vx)\vect{y} \leq \epsilon_2 \vect{y}^\mathrm{T}\vect{y}, \vx \in \Omega,~ \forall\vect{y} \in \mathbb{R}^{\textit{nd}}
\end{equation}
\subsection{Govering equations for subsurface flow}

The governing equation for subsurface flow is given by
\begin{align}
\frac{\partial}{\partial t}\left(\rho \phi \right) - \divergence{\left(\frac{\rho k}{\mu} \gradient{p}\right)} = Q_m,
\end{align}
where $\phi$ is the porosity (unitless), $\rho$ is the mass density (kg/m$^3$), $\mu$ is the dynamic viscosity (Pa-s), $k$ is the permeability (m$^2$), $p$ is the pressure (Pa), $Q_m$ is the volumetric flow rate (kg/$m^3$/s). Assuming $\phi$ is constant and that the spatial gradient of density is small, the above equation reduces to
\begin{align}
\rho\phi \beta \frac{\partial p}{\partial t} -   \frac{1}{g}\divergence{\left(K \gradient{p}\right)} = Q_m,
\end{align}
where $\beta$ is water compressibility ($\frac{1}{\rho}\frac{\partial \rho}{\partial p}$, Pa$^{-1}$) and $K$ is the hydraulic conductivity (m/s). Here permeability is connected to hydraulic conductivity by $k=K\mu / (g\rho )$, and $g$ is the acceleration due to gravity ($m/s^2$). We shall denote the pressure field by $c(\vect{x})$. Let us consider the transient flow in heterogeneous porous media governed by the following diffusion equation and boundary/initial conditions
\begin{equation}
\label{equ:tranequ}
\begin{aligned}
\dot{u}(\vx,t) &= \divergence[\vect{D}(\vx)\gradient{u(\vx,t)}] +b(\vx,t),~\vx \in \Omega, ~t\in[0,T] \\
u(\vx,t) &= u^{p}(\vx,t), ~\textrm{on}~ \Gamma^{D} \\
-\hat{\vect{n}} \cdot \vect{D}(\vx)\gradient{u(\vx,t)} &= q^{p}(\vx,t), ~\textrm{on}~ \Gamma^{N} \\
u(\vx,0) &= u_0(\vx), ~\textrm{in}~ \Omega,
\end{aligned}
\end{equation}
where $b(\vx,t)$ is the volumetric source or sink, $u^{p}(\vx,t),q^{p}(\vx,t)$ are prescribed pressure and flux respectively. $D(x)$ is the scaled diffusivity as D(x) = $K(\vx)/(g\rho \phi \beta)$. For uniqueness, we assume $\Gamma^{D}$ is not empty. This initial-boundary-value-problem(IBVP) is a second-order parabolic partial differential equation(PDE). Let $L=\frac{\partial}{\partial t}-\divergence[{\vect{D(\vx)}(\gradient)}]$ denote the operator in Eq. \ref{equ:tranequ}, it is worthwhile to point out that the adjoint operator is $L^*=-\frac{\partial}{\partial t}-\divergence[{\vect{D(\vx)}(\gradient)}]$, where time runs backwards. The adjoint problem to Eq. \ref{equ:tranequ} involves adjoint boundary conditions, which is often non-trivial to formulate, especially when irregular boundary configuration is involved.
\subsubsection{Maximum principle}
The maximum principle of a transient diffusion equation asserts that the maximum can occur only on the boundary of the domain or in the initial condition if $b(\vx, t) \leq 0$ and $\Gamma^{D}=\partial\Omega$. Mathematically, a solution to equations (2.18a)–(2.18a)
will satisfy:

\subsection{PDE-constrained optimization }
Determining parameters of a partial differential equations(PDE) model is often formulated as a PDE-constrained optimization problem where the field values mathch observations. This is also referred as inverse problem. 
Such problems take the form,
\begin{equation}
\label{eq:opti}
\begin{aligned}
& \underset{x}{\text{min}}
& &J(u,p) \\
& \text{s.t.} & &  F(u,p) = 0, \\ \nonumber
\end{aligned}
\end{equation}
where $u$, $p$, $J(u,p)$ and $F(u,p)$ are field value, parameters in PDE, objective function and PDE induced constraints. From a optimization point of view, it is required that $u$ be feasible at every step in $p$ when $J(u,p)$ converges to a minimizer. The necessary ingredients of a capable optimization solver for Eq. \ref{eq:opti} should: 1)be able to solve $F(u,p)=0$(PDE or forward problem solver); 2) evaluate $J(u,p)$; 3) provide the gradient $\text{d}_pJ$. 
Among those problems, time-dependent ones arise wide attentions for such a reason that forward problems are often treated by the method-of-line which induces a system of ODE. The adjoint equation to the probelm is also an ODE, which means that they both can be solved by the same standard ODE integrators. The adjoint method of time-dependent problem comes in the form,
\begin{equation}
\label{eq:td-adjoint}
\begin{aligned}
& \underset{x}{\text{min}}
& &\Psi_i(u_0,p) = \Phi_i(u_T,p) + \int_{0}^{T} r_i(t,u(t),p)dt,~i=1,...,n_{\text{obj}}\\
& \text{s.t.} & &  F(t,u,\dot{u},p) = 0,~0\leq t \leq T\\
&  & & u(0) = u_0(p) \\ 
\end{aligned}
\end{equation}
The ith total derivative(gradient) is denoted as,
\begin{equation}
\label{eq:totalderi}
\begin{aligned}
\text{d}_p \Psi_i(u,p) = \text{d}_p\Phi_i(u_T,p) +\int_0^T[\partial_ur_i\text{d}_pu+\partial_pr_i]dt \\
\end{aligned}
\end{equation}
The corresponding Lagrangian of \ref{eq:td-adjoint} can be written as
\begin{equation}
\mathcal{L}_i=\Phi_i(u_T,p)+\int_0^T[r_i+\nu_i^T F(t,u,\dot{u},p)]dt + \mu_i^T[u(0)-u_0(p)],
\end{equation}
where $\nu_i$ and $\mu_i$ are vectors of Lagrangian multipliers as function of time. They are also named by adjoint vectors. Since only equality constraints are involved, we are free to set values of $\nu_i$ and $\mu_i$. Also note that $d_p\mathcal{L}_i = d_p\Psi_i$, the total derivative is,
\begin{equation}
\begin{aligned}
\label{eq:toalderi2}
\text{d}\mathcal{L}_i &= \text{d}_p\Phi_i(u_T,p)+\int_0^T[\partial_u r_i\text{d}_pu+\partial_pr_i+\nu_i^T(\partial_uF\text{d}_pu+\partial_{\dot{u}}F \text{d}_p\dot{u}\partial_pF)]dt \\ &+\mu_i^T[\text{d}_pu(0)-\partial_pu_0(p) ] \\
 &= \text{d}_p\Phi_i(u_T,p)+\int_0^T\{[\partial_ur_i+\nu_i^T(\partial_uF-\text{d}_t\partial_{\dot{u}}F)-\dot{\nu_i}^T\partial_{\dot{u}}F]\text{d}_pu+\partial_pr_i\\
 &+\nu_i^T\partial_pF\}dt + \nu_i^T\partial_{\dot{u}}F\text{d}_pu\rvert_T + (\mu_i^T-\nu_i^T \partial_{\dot{u}}F)\rvert_0 \text{d}_pu(0) -\mu_i^T\partial_pu_0(p),
\end{aligned}
\end{equation}
where integration by part is used. The term $\text{d}_pu\vert_T$ is non-trivial to obtain, thus we set $\nu_i\vert_T=0$ to make the whole term vanish. By setting $\mu_i^T\rvert_0=\nu_i^T \partial_{\dot{u}}F\rvert_0$, evaluation of term $\text{d}_pu(0)$ is avoided. Recursively, we can avoid computing $\text{d}_pu$ for all $t > 0$ by setting
\begin{equation}
\partial_ur_i + \nu_i^T(\partial_uF-\text{d}_t\partial_{\dot{u}}F)-\dot{\nu_i}^T\partial_{\dot{u}}F = 0
\end{equation}
The following algorithm describes how $\text{d}_p\Psi_i$ is computed:\\
\begin{algorithm}[H]
 \KwData{$u_0(p)$}
 \KwResult{$\text{d}_p\Psi_i$}
  1)Forward step: integrating $F(t,u,\dot{u},p) = 0$ over time from $t = 0$ to $T$ of $u$ with initial condition  $u(0) = u_0(p)$\\
  2) Adjoint step: integrating $\partial_ur_i + \nu_i^T(\partial_uF-\text{d}_t\partial_{\dot{u}}F)-\dot{\nu_i}^T\partial_{\dot{u}}F = 0$ over time from $t = T$ to $0$ of $\nu_i$ with initial condition $\nu_i \rvert_T =0$\\
  3) $\text{d}_p\Psi_i=\text{d}_p\Phi_i(u_T,p)+\int_0^T(\partial_pr_i+\nu_i^T\partial_pF)dt+\nu_i^T\partial_{\dot{u}}F\rvert_0\partial_pu_0(p)$ .
 \caption{Computing gradient of objective function $\Psi_i$}
\end{algorithm}
The output $\text{d}_p\Psi_i$ is the Jacobian matrix which is associated with the sensitivity on $p$. It only takes 1 forward and 1 adjoint(inverse) run, the Jacobian is yielded. As a compare, differentiation based approach needs to take $\text{dim}(p)$ forward runs. The advantages get siginificant when $n_{\text{obj}} \gg \text{dim}(p)$.
\subsubsection{Discrete adjoint sensitivity analysis}
There are several ways to solve Eq. \ref{equ:tranequ}, here the method of line approach is adopted, which resulting a system of ordinary differential equations(ODE) as,
\begin{equation}
\label{equ:ODE}
\mathcal{M}\dot{u}(t)=f(t,u(t)), u(0)=\beta
\end{equation}
where $u(t)$ is the spatial discretization of flow field $u(\vx,t)$. $\mathcal{M}$ is the mass matrix which is usually symmetric-positive definite. Here assume $\mathcal{M}$ is identity for brevity of notations. The right-hand-side $f(t,u)$ involves the contribution from the parameters of model(permeability distribution $\vect{D}(\vx)$). Let us consider a simples t forward integration scheme, backward Euler, for \ref{equ:ODE} as
\begin{equation}
\label{equ:BackEu}
\mathcal{M}u_{n+1} =\mathcal{M}u_{n}+\dt f(t_{n+1},u_{n+1})
\end{equation}
Now define the sensitivity variable as $\vect{S}_{l,n}=\partial u_n / \partial p_l$, where $p_l$ means the $p$th parameters in the model. The sensitivity equation corresponding to $p_l$ is immediately obtained after pluging $\vect{S}_{l,n}$ into Eq. \ref{equ:BackEu},
\begin{equation}
\label{equ:sensiEqu}
\mathcal{M}\vect{S}_{l,n} = \mathcal{M}\vect{S}_{l,n}+\dt (f_u(t_{n+1},u_{n+1})\vect{S}_{l,n+1}+f_{p}(t_{n+1},u_{n+1}))
\end{equation}
where $f_u$ and $f_p$ are Jacobian matrices. As we can see that the trajactory of $\vect{S}_{l,n}$ follows a similar trajactory with model's state variable in the forward process. To be general, use $u_{n+1}=\mathcal{N}_n(u_n),n=0,...,N-1$ to denote any one-step integration scheme. In our implementation , the objective function $\Phi$ is chosen to only involve the terminal term under the effect of parameters $p$ as
\begin{equation}
\Phi=\phi(u(T);p).
\end{equation}
The constraints of the optimization problem are chosen to be the discretized PDE at each time step. Therefore, the Lagrangian is written as
\begin{equation}
\mathcal{L}=\phi(u_N) - \nu_0^T(u_0-\beta)\sum_{n=0}^{N-1}\nu_{n+1}^{T}(u_{n+1}-\mathcal{N}(u_n))
\end{equation}
,where $\nu_0,...,\nu_N$ are Lagrange multipliers. We use $\phi(u_N)$ to approximate $\phi(u(T))$. Differentiating this function with respect to $p$ yields
\begin{equation}
\frac{\partial \mathcal{L}}{\partial p}=\nu_0^T\frac{\partial \beta}{\partial p}-(\frac{\partial \phi(u_N)}{\partial u}-\nu_N^T)\frac{\partial u_N}{\partial p}- \sum_{n=0}^{N-1}(\nu_{n}^T- \nu_{n+1}^T \frac{\partial \mathcal{N}(u_n)}{\partial u})\frac{\partial u_n}{\partial p}
\end{equation}
Let $\partial \mathcal{L}/ \partial p=0$ and define 
\begin{equation}
\label{equ:recu}
\nu_N^T=\frac{\partial \phi(u_N)}{\partial u}~,\nu_n^T=\nu_{n+1}^T\frac{\partial \mathcal{N}(u_n)}{\partial u},~ n=N-1,...,0
\end{equation} 
The gradient of target objective function is 
\begin{equation}
\nabla_p\phi=(\frac{\partial \beta}{\partial p})^T\nu_0
\end{equation}
Now treat $\mathcal{N}(u)$ as a implicit function and use backward Euler as example. Take derivative of $u$ in Eq. \ref{equ:BackEu}, we get
\begin{equation}
\frac{\partial u_{n+1}}{\partial u} = \frac{\partial u_{n}}{\partial u}+\dt f_u(t_{n+1},u_{n+1})\frac{\partial u_{n+1}}{\partial u}=\frac{\partial \mathcal{N}(u_n) }{\partial u} \frac{\partial u_{n}}{\partial u},
\end{equation}

Combining with Eq. \ref{equ:recu}, the discrete adjoint equation is formulated as,
\begin{equation}
\nu_n^T=\nu_{n+1}^T+\dt \nu_n^T f_u(t_{n+1},u_{n+1}).
\end{equation}

\section{Numerical Implementation}
\subsubsection{PETSc and TAO}
We leverage on scientific libraries such as PETSc and TAO to implement the large-scale inversion's computation. PETSc is a suite of data structures and routines
for the scalable (parallel) solution of scientific applications modeled by PDEs, implementing MPI standard and widely used in parallel finite element codes development. It also provides interfaces to several other libraries such as Metis/ParMETIS and HDF5 for mesh partitioning and binary data format handling
respectively. To solve the large-scale optimization problem, another important feature with PETSc, TAO, is employed. Our non-negative methodology will use the  Bounded Limited-Memory Variable-Metric(BLMVM) solver available in TAO to approximate the Hessian, and this is efficient in large-scale context. Other optization algorithm like Conjugate Gradient(CG) and Limited-Memory Variable-Metric(LMVM) will be compared in the convergence and memory consumption. Further details regarding the implementation of these various methods may be found in  and the references within.
\subsubsection{Weak formulation}
Continuous Galerkin approach is adopted for the FE setup. The trial and test function spaces are chosen to be 
\begin{equation}
\begin{aligned}
\mathcal{U}&:=\{c(\vx ) \in H_{\Gamma ^D}^1(\Omega) \rvert c(\vx) = c^p(\vx)~ \text{on} ~\Gamma ^D\} \\
\mathcal{W}&:=\{w(\vx ) \in H_{0}^1(\Omega) \rvert w(\vx) = 0~ \text{on} ~\Gamma ^D\}
\end{aligned}
\end{equation}
 The weak form for Eq. \ref{equ:tranequ} reads: find $c(\vx)\in \mathcal{U}$, such that 
\begin{equation}
\mathcal{B}(w;c) = L(w),~\forall w \in \mathcal{W}
\end{equation}
where the bilinear form and linear functional are, respectively, defined as
\begin{equation}
\label{eq:weakform}
\begin{aligned}
\mathcal{B}(w(\vx);c(\vx,t_n)) &:= \frac{1}{\Delta t}\int_{\Omega} w(\vx)c(\vx,t_n)+ \Delta t\gradient[w(\vx)] \cdot \vect{D}(\vx)\gradient[c(\vx,t_n)]\text{d}\Omega \\
L(w(\vx)) &:= \frac{1}{\Delta t}\{\int_{\Omega}w(\vx)[b(\vx,t_n)dt+c(\vx,t_{n-1})]\text{d}\Omega +\int_{\Gamma^N}w(\vx)q^p(\vx)\text{d}\Gamma  \}
\end{aligned}
\end{equation}

The assembly of mass/stiffness matrix, Gaussian quadrature and other routines are implemented in-house while the parallel matrix/vector operations are interfaced with in PETSc's build-in. First, following the FE model outlined in [19], we consider the weak form that depends on fields and
gradients. The residual evaluation can be expressed as:
\begin{equation}
w^Tr(c)~\int_{\Omega^e}[w\cdot\mathcal{F}_0(c,\gradient c)+\gradient w \cdot \mathcal{F}_1(c,\gradient c)]\text{d}\Omega = 0,
\end{equation}
where $\mathcal{F}_0(c,\gradient c)$ and $\mathcal{F}_1(c,\gradient c)$ are point-wise functions that capture the physics. This framework decouples the problem specification from the mesh and degree of freedom traversal which easy the implementation on distributed memory machines. The discretization of the residual is written as:  
\begin{equation}
\label{eq:disres}
r(c) = \mathbf{A}_{\text{e=1}}^{\text{Nele}}\begin{bmatrix}N^T & B^T\end{bmatrix}W
\begin{bmatrix}
\mathcal{F}_0(c_q,\gradient c_q)\\\mathcal{F}_1(c_q,\gradient c_q)
\end{bmatrix}
\end{equation}
where $ \mathbf{A}$ represents the assembly operator, $N$ and $B$ are matrix forms of basis functions over quadrature points, diagonal matrix $W$ is the quadurature weights, and $c_q$ is the field value at quadrature point $q$. Mapping back to Eq. \ref{eq:weakform},
\begin{equation}
\mathcal{F}_0 = \frac{1}{\Delta t}[c^n_q(\vx)  - c^{n-1}_q(\vx)]- b^n(\vx),~\mathcal{F}_1 =  \vect{D}(\vx)\gradient c^n_q
\end{equation}
here the superscript $n$ denotes for time step and also assume $q^p(\vx)=0$ for simplicity. 
Naturally, the Jacobian is the derivatives of Eq.\ref{eq:disres} as,
\begin{equation}
\begin{aligned}
J(c) &= \mathbf{A}_{\text{e=1}}^{\text{Nele}}\begin{bmatrix}N^T & B^T\end{bmatrix}W\begin{bmatrix}
\mathcal{F}_{0,0} &\mathcal{F}_{0,1}\\
\mathcal{F}_{1,0} &\mathcal{F}_{1,1}
\end{bmatrix}\begin{bmatrix}
N\\
B
\end{bmatrix}
\\
[\mathcal{F}_{i,j}]&=\begin{bmatrix}
\partial_{c}\mathcal{F}_{0} &\partial_{\gradient c} \mathcal{F}_{0}\\
\partial_{c} \mathcal{F}_{1} &\partial_{\gradient c} \mathcal{F}_{1}
\end{bmatrix}(c_q,\gradient c_q).
\end{aligned}
\end{equation}
The point wise functions are
\begin{equation}
\mathcal{F}_{0,0}  = \frac{1}{\Delta t},~ \mathcal{F}_{0,1} = 0,~\mathcal{F}_{1,0} = 0,~\mathcal{F}_{1,1} = \vect{D}(\vx)
\end{equation}

\subsubsection{Parallel finite element assembly}
In each optimization step, one forward and adjoint run are conducted and each run is a solution of time-dependent problem. The PETSc interface for solving time dependent problems assumes the problem is written in the form
\begin{equation}
F(t,u,\dot{u}) = G(t,u),~ u(0)= u_0.
\end{equation}
User has to provide how to evaluate the residual and Jacobian from $F(t,u,\dot{u})$ using interface functions "TSSetIFunction" and "TSSetIJacobian". Take backward Euler scheme applied to $F(t,u,\dot{u},p) = 0 $ as example, the time derivate $\dot{u} = (u_n-u_{n-1})/\Delta t$, it results in the Jacobian $\partial_{u^n}F= I/\Delta t+\partial_uF(t,u,\dot{u},p)$. As a result, evalulation of the Jacobian for each forward/adjoint run is required since it is a function of $p$. But within one forward(or adjoint)run, it just has been computed once if fixed time step is assumed. Apart from matrix systems solution by Krylov method, matrices assembly is another bottle-neck when going to large scale.

\begin{figure}[H]
	\centering
	\includegraphics[scale=0.45]{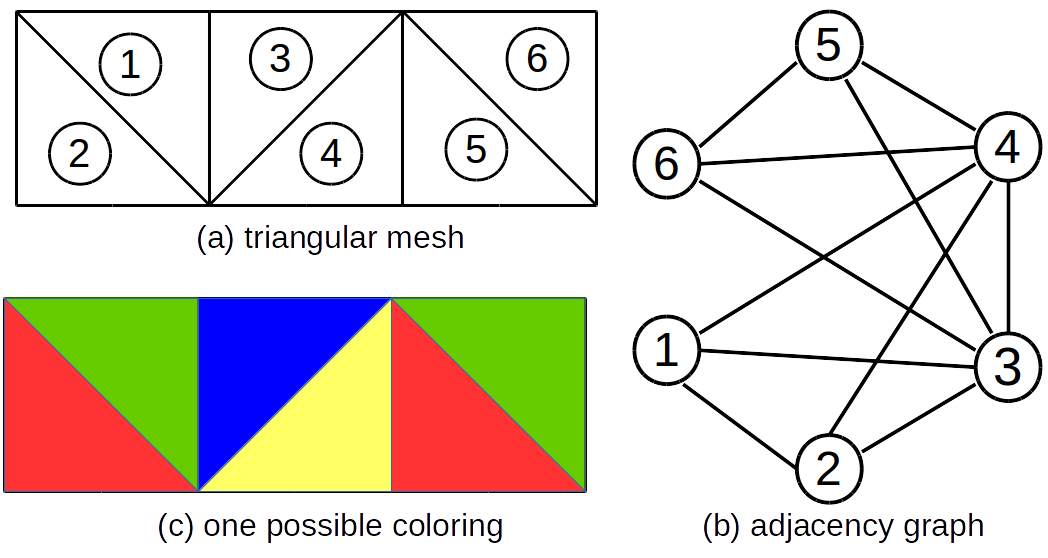}
	\caption{Example of graph coloring of triangular mesh}
	\label{fig:adjexp}
\end{figure}

This paper considers a hybrid framework of parallelism on both shared-memory(OpenMP) and distributed-memory(MPI) level. In dealing with shared memory machines, the assembly of stiffness matrices in FE will encounter race condition if two adjacent elements are assembled at the same time by two threads. With the help of graph coloring, the elements can be assembled one color at a time, thus preventing race condition. In order to do the coloring, the indices of neighboring elements are necessary ,which can be readily obtained from the adjacency graph of the mesh. Take the triangular mesh in Fig \ref{fig:adjexp} as example, the corresponding graph and one possible coloring(4-colored) are shown.
Since the test and trial functions are nodal based, two elements are considered to be connected once they share at least one node. Elements in the same color now are safe to be assembled by different threads.

\section{Results and Performance}

\subsection{2D Verification}

\subsubsection{Hetergeneous diffusion in 2D geometry}

\begin{figure}[H]
\centering
\subfigure[Step 2]{\includegraphics[scale=0.3]{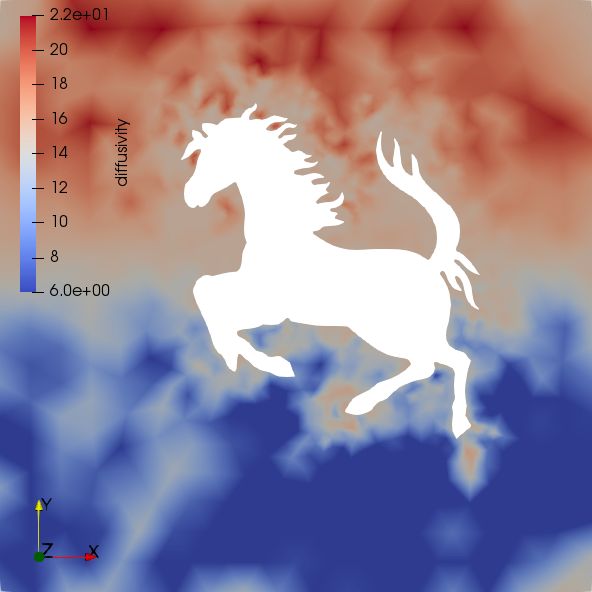}}
\subfigure[Step 10]{\includegraphics[scale=0.3]{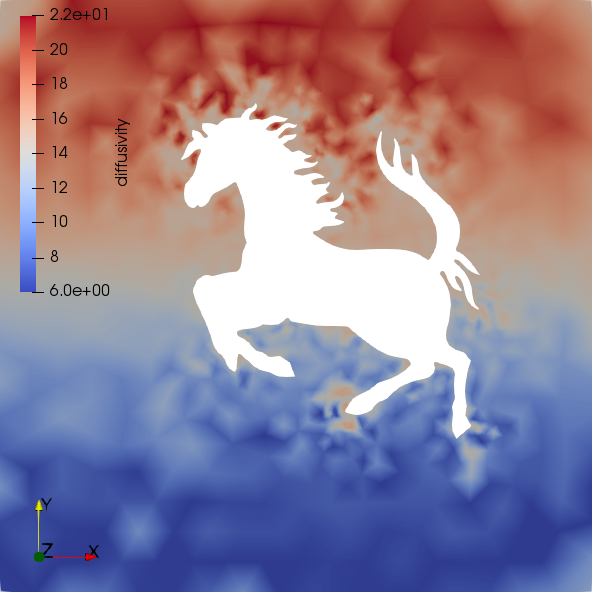}}\\
\subfigure[Step 30]{\includegraphics[scale=0.3]{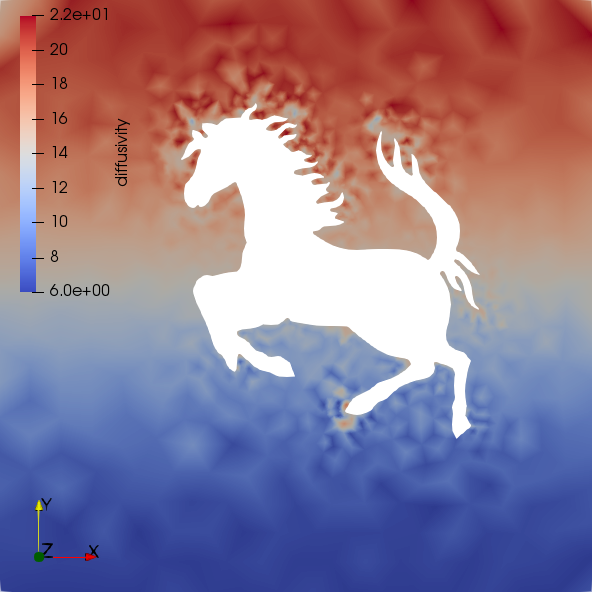}}
\subfigure[Step 60]{\includegraphics[scale=0.3]{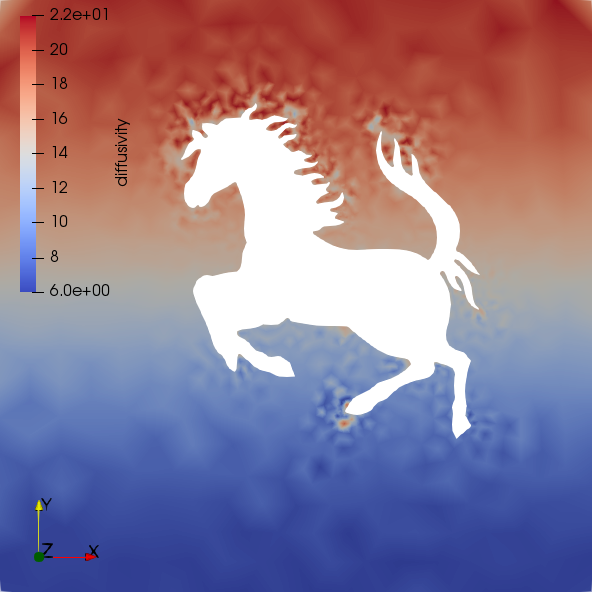}}\\
\subfigure[Step 100]{\includegraphics[scale=0.3]{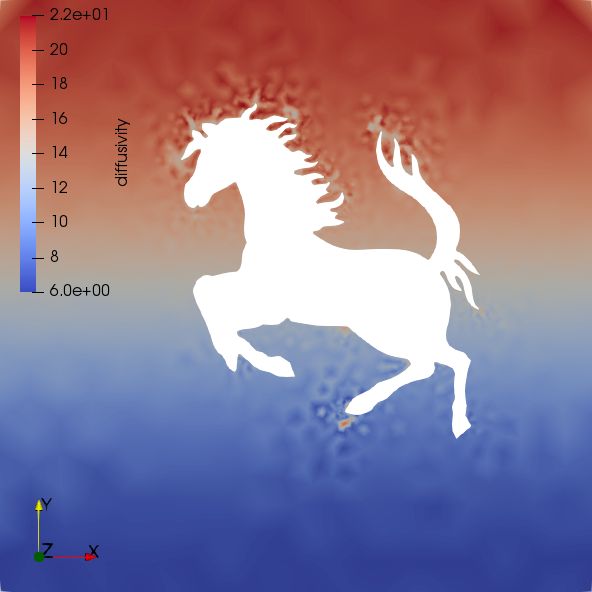}}
\subfigure[True distribution]{\includegraphics[scale=0.3]{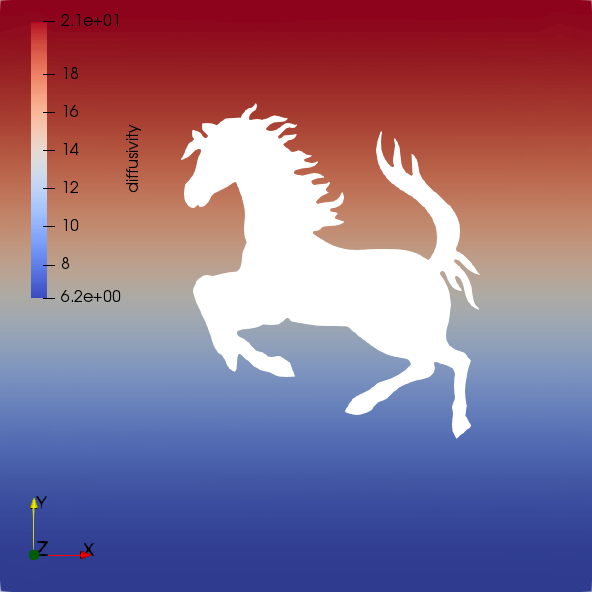}}\\
\caption{put a color map figure here! Diffusivity distribution after optimization steps}	\label{fig:2dhorse-out}
\end{figure}

Convergence with tao types
\begin{figure}[H]
\centering
\subfigure[]{\includegraphics[scale=0.5]{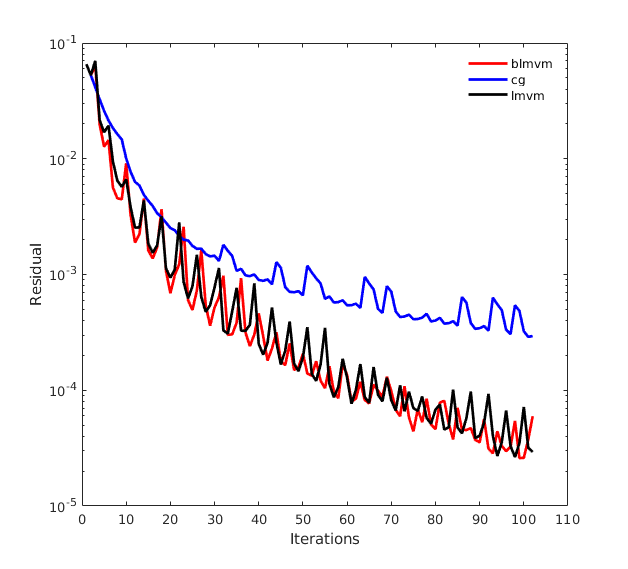}}
\caption{Convergence of different optimization solvers}	\label{fig:2dhorse-out}
\end{figure}

\subsubsection{Hetergeneous diffusion in a unit cube with spherical holes}
Let the computational domain
be a unit cube with two spherical holes of radius 0.2 and 0.35. The concentration on the outer boundary is taken to be zero and the concentration on the interior boundary is taken to be unity.
The volumetric source is taken as zero (i.e., f (x) = 0). The velocity vector field for this problem is
chosen to be
\begin{figure}[H]
\centering
\subfigure[Location of the hole]{\includegraphics[scale=0.25]{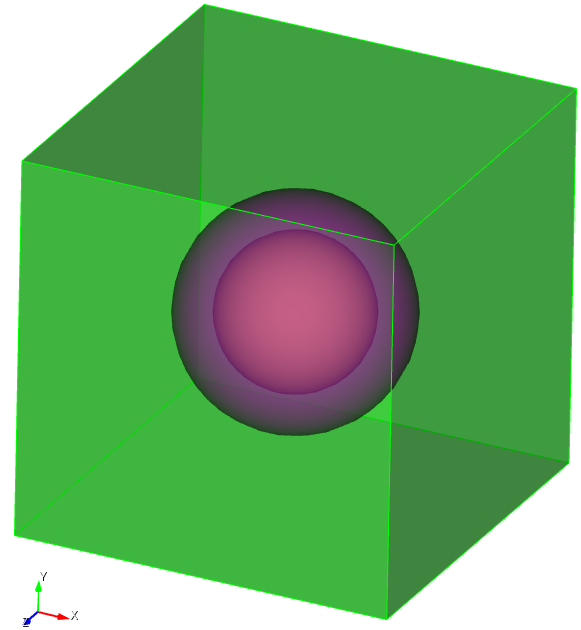}}
\subfigure[Mesh type A]{\includegraphics[scale=0.25]{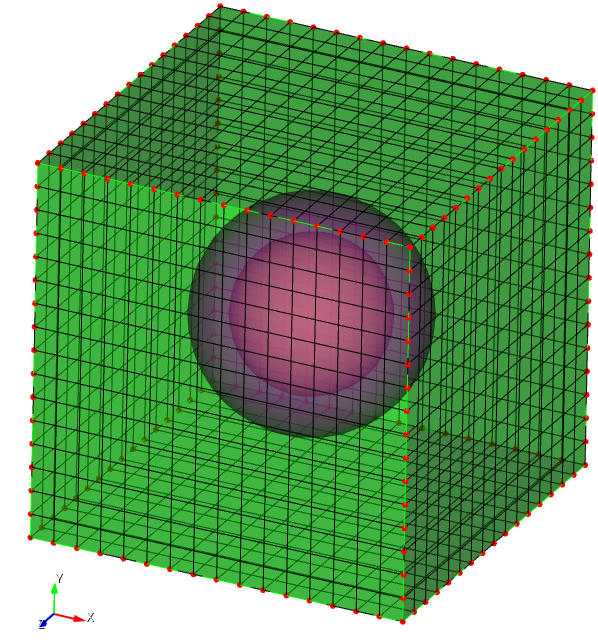}}
\subfigure[Mesh type B]{\includegraphics[scale=0.25]{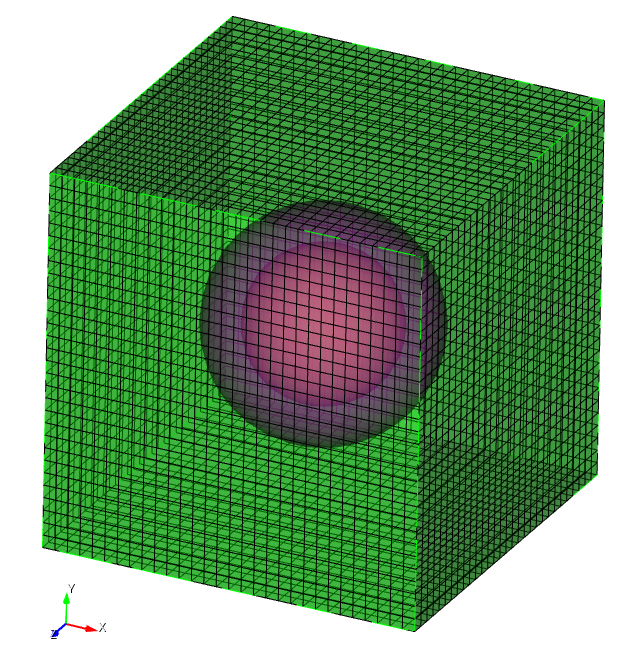}}
\caption{Cube with a hole: pictorial description and the associated grids}
\end{figure}

The estimation of diffusivity for mesh type B after three optimization steps are shown below.
\begin{figure}[H]
\centering
\subfigure[Step 2]{\includegraphics[scale=0.25]{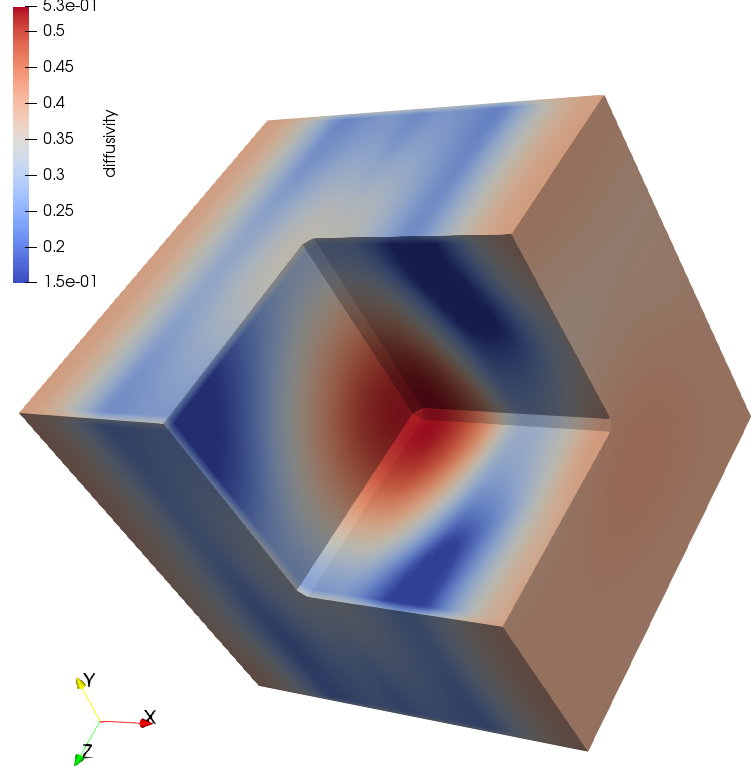}}
\subfigure[Step 10]{\includegraphics[scale=0.25]{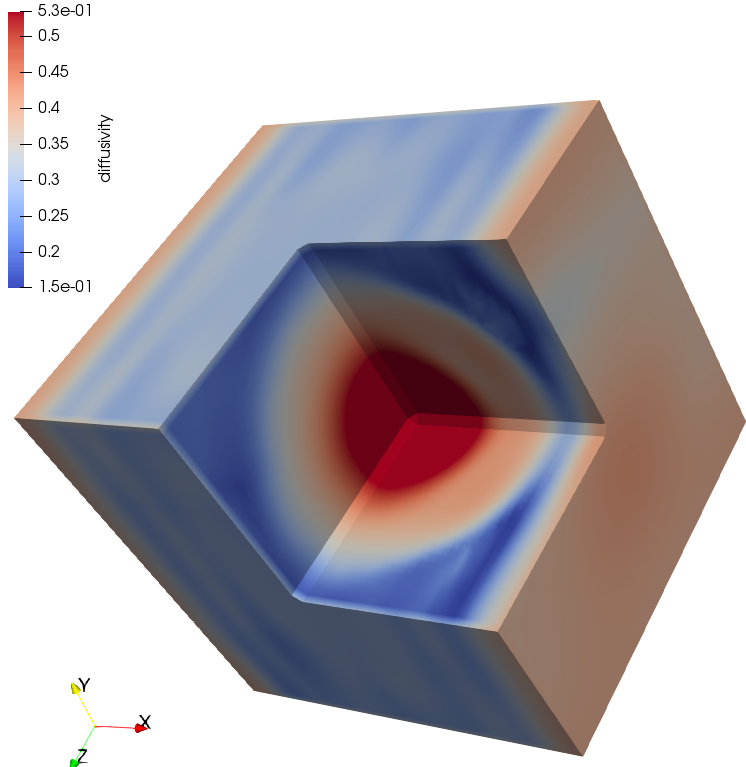}}\\
\subfigure[Step 100]{\includegraphics[scale=0.25]{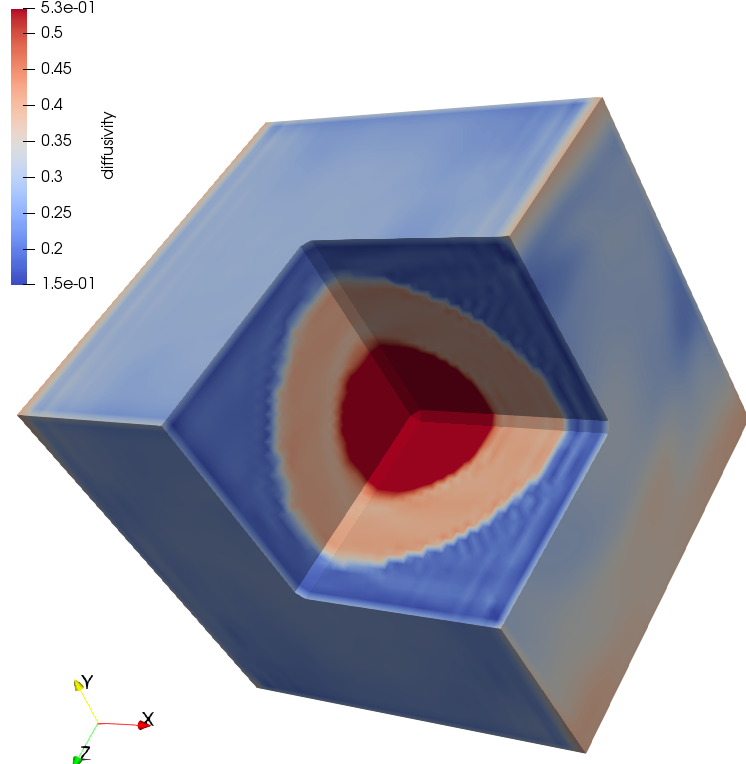}}
\subfigure[True diffusivity]{\includegraphics[scale=0.25]{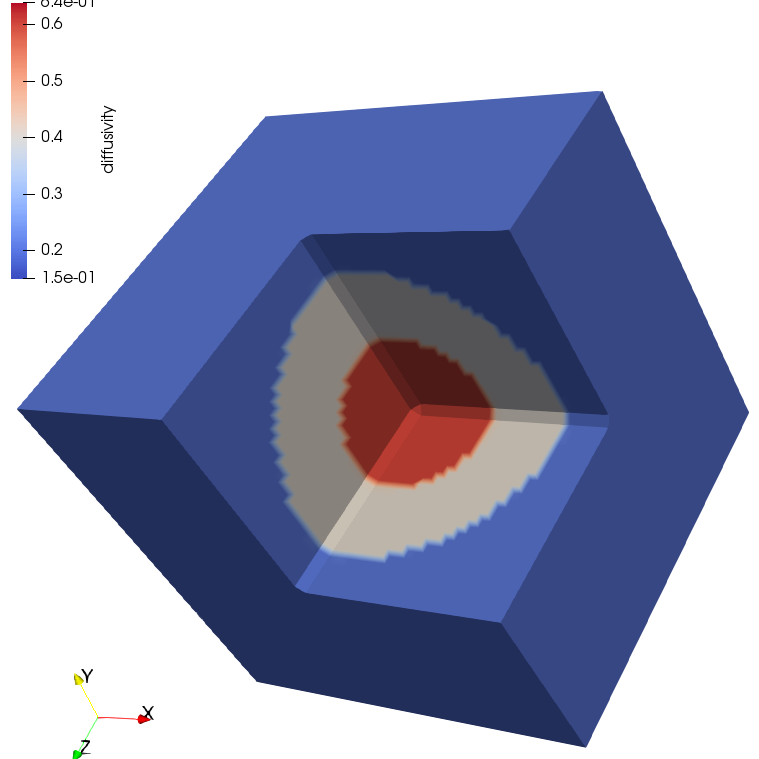}}
\caption{Cube with a hole: inversion result and true solution}
\end{figure}

\section{Parallel Performance}
\subsection{Strong scaling performance of forward run}
1million hex grid forward run
\begin{figure}[H]
\centering
\subfigure[]{\includegraphics[scale=0.5]{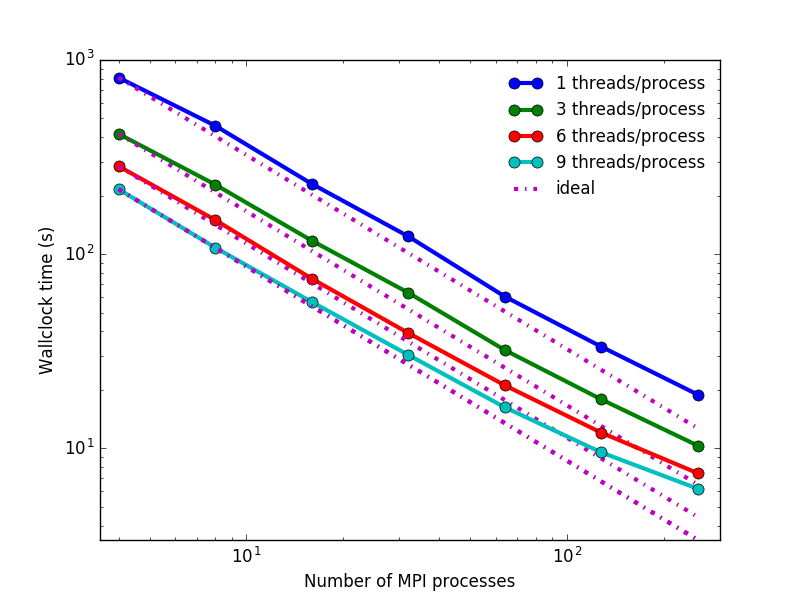}}
\caption{Strong scaling}
\end{figure}

\subsubsection{Scaling performance of multi-threading}
The multi-threading is implemented with OpenMP on multi-core CPUs.
\begin{figure}[H]
\centering
\subfigure[]{\includegraphics[scale=0.5]{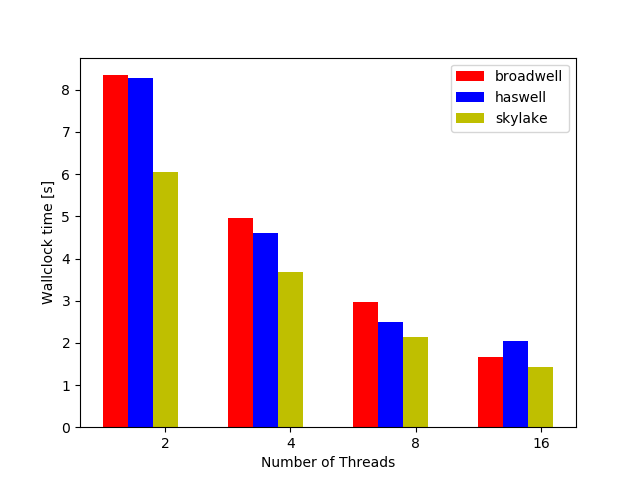}}
\caption{OpenMP scaling}
\end{figure}

\subsubsection{Scaling performance of MPI+OpenMP}
\begin{figure}[H]
\centering
\subfigure[]{\includegraphics[scale=0.5]{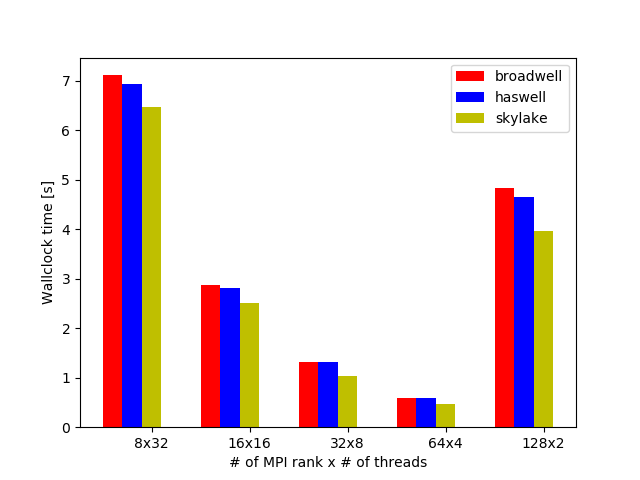}}
\caption{MPI+OpenMP scaling}
\end{figure}

\subsection{Performance modeling}
Since the inversion process involves both forward and backward runs, as well as optimization steps, there will be fraction of the code that is not amenable to parallelization. Here we employ Amdahl's law and Gustafson's law to model strong and weak scaling respectively.   
\subsection{Strong scaling model}
Amdahl’s law can be formulated as follows
\begin{equation}
\text{Speed-up}=\frac{1}{s+\frac{1-s}{N}}
\end{equation}
where s is the proportion of execution time spent on the serial part and N is the number of processors. Amdahl’s law states that, for a fixed problem, the upper limit of speedup is determined by the serial fraction of the code. 
Here we tested on three types of mesh with unknown sizes being 0.25, 1 and 4 million. The results and fitted model are shown below.

\begin{figure}[H]
\centering
\subfigure[Speed-up and efficiency]{\includegraphics[scale=0.4]{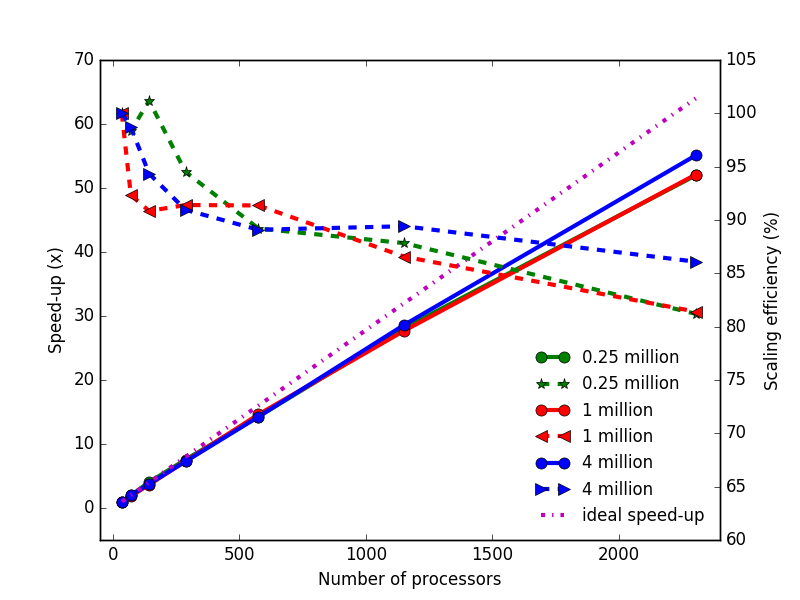}}
\subfigure[Fitted model]{\includegraphics[scale=0.32]{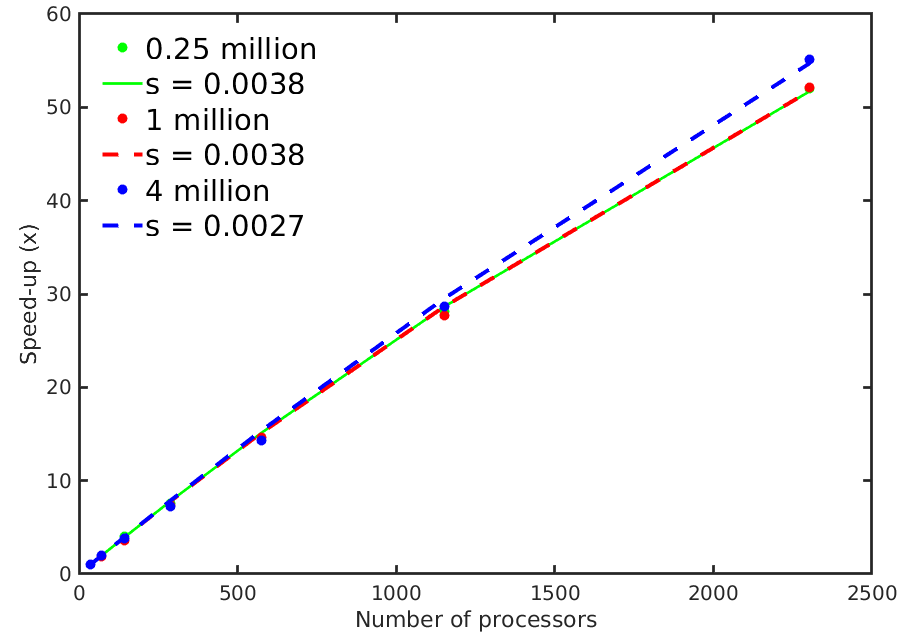}}
\caption{Strong scaling model}
\end{figure}
As dipicted in the figure, the fraction of serial part of the code decreases with the increasing of problem size. Thus we expect better strong scaling performance for large problem.

\subsection{Weak scaling model}
The sizes of problems scale with the amount of available resources in real applications. A more reasonable choice is to use small amounts of resources for small problems and larger quantities of resources for big problems. Amdahl’s law gives the upper limit of speedup for a problem of fixed size. For measuring the weak scaling, where the scaled speedup is calculated based on the amount of work done for a scaled problem size (in contrast to Amdahl’s law which focuses on fixed problem size), Gustafson’s law is a more wise choice. It is based on the approximations that the parallel part scales linearly with the amount of resources, and that the serial part does not increase with respect to the size of the problem. It provides the formula for scaled speedup as:
\begin{equation}
\text{Scaled speed-up}=s+(1-s)N
\end{equation}
, where $s$ and $N$ has the same meaning as in Amdahl’s law. 
Here we fix the number of cells per processor and increase the number of processors. The results and fitted model are shown below.
\begin{figure}[H]
\centering
\subfigure[Scaled speed-up and efficiency]{\includegraphics[scale=0.4]{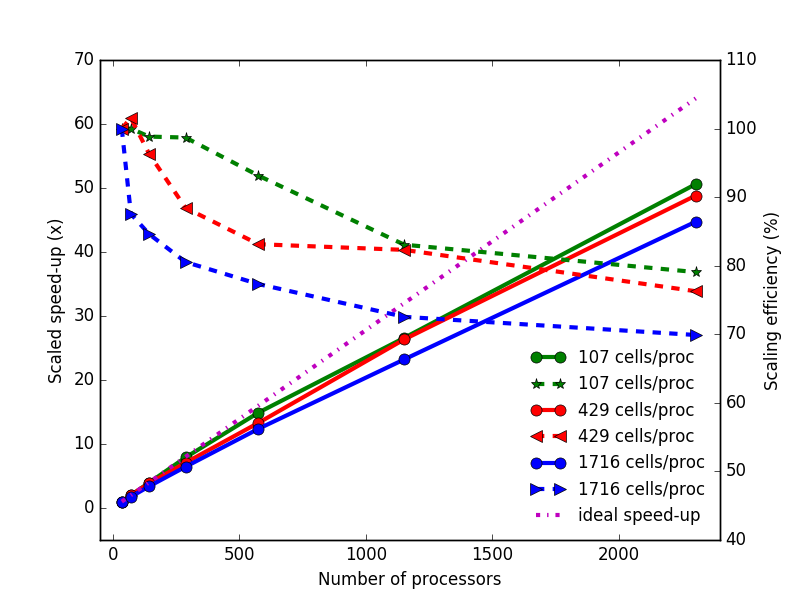}}
\subfigure[Fitted model]{\includegraphics[scale=0.32]{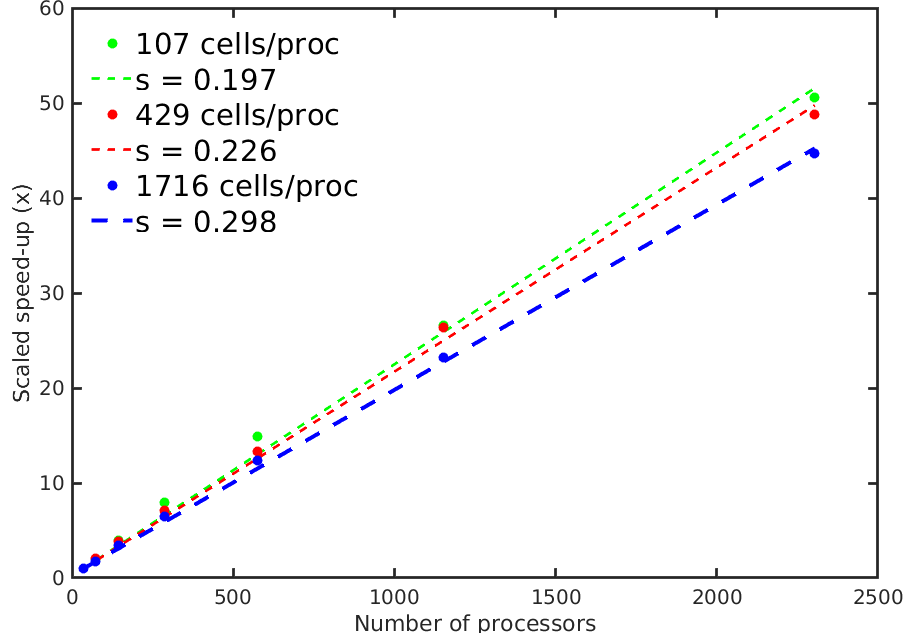}}
\caption{Weak scaling model}
\end{figure}

We observe that, as incresing the workload per processor, the weak scaling gets worse, together with the proportion of serial part increases. This can be explained by the fact that the optimizer, which is the major serial part, takes more efforts to find next optimization direction.  Also notice that the  discrepancy in s between strong/weak scaling modeling. This is attributed to the approximations in the laws — the serial fraction is assumed to remain constant, and the parallel part is assumed to be speed up in proportion to the number of processors. In practice, the overhead of parallelization may also increase with the job size (e.g. from the scheduling of threads), and in this case it is understandable that the weak scaling model gives a larger serial fraction s.
\subsection{Real-world Problem}
\begin{figure}[H]
\centering
\subfigure[Pressure at day 1]{\includegraphics[scale=0.25]{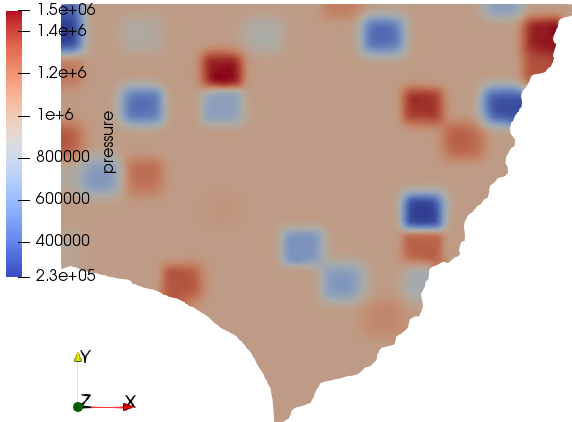}}
\subfigure[Pressure at day 150]{\includegraphics[scale=0.25]{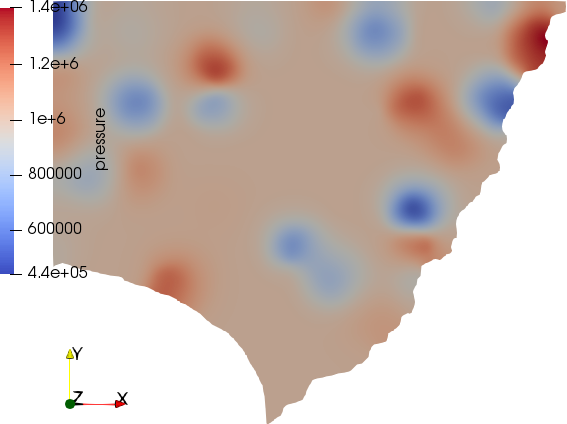}}
\label{fig:real-data}
\caption{Pressure observed in 2D}
\end{figure}

We consider a real world problem of subsurface flow in this section. The parameters of simulation domain is described in the table below. 
In the first case, a 2D model is considered. The inversion run is carried out based on the observation of pressure collected at day 1 and day 150. In this simulation, only the flows on x-y plane are considered. The initial pressure is only collected at 25 locations and the background pressure is assumed to be $1.0 \times 10^{6}$ Pa. The pressure at day 1 and 150 are shown in Fig. \ref{fig:real-data}. 

\begin{figure}[H]
\centering
\subfigure[Iteration 50]{\includegraphics[scale=0.3]{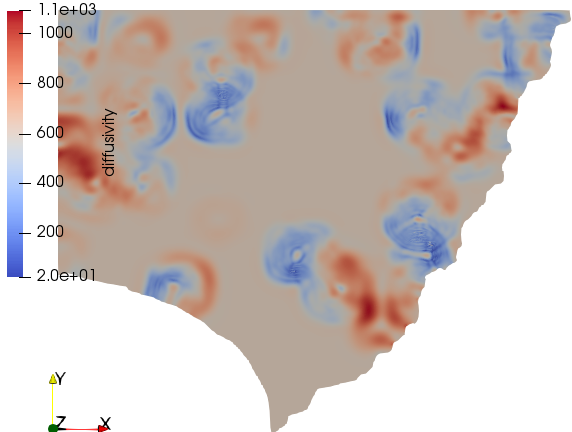}}
\subfigure[Iteration 110]{\includegraphics[scale=0.3]{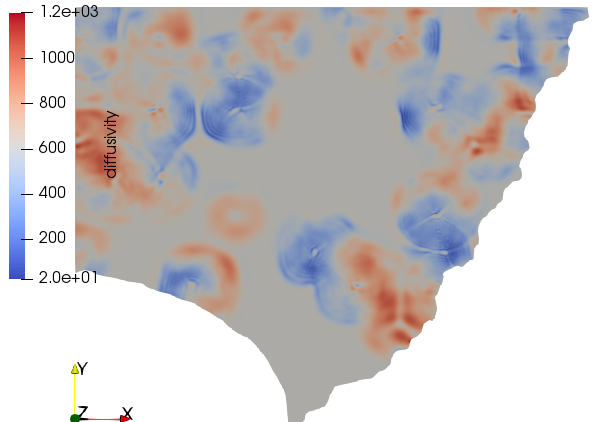}}\\
d\subfigure[True diffusivity]{\includegraphics[scale=0.3]{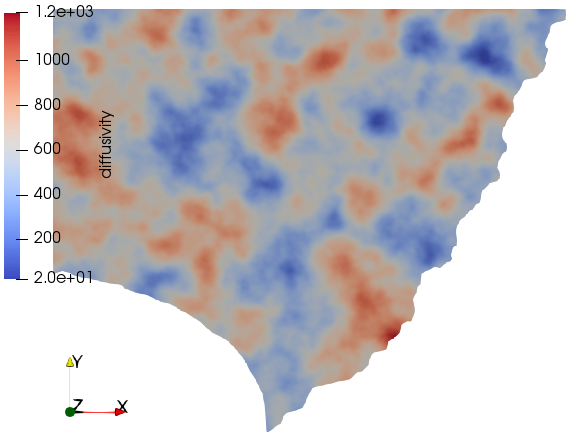}}
dim\subfigure[Convergence]{\includegraphics[scale=0.37]{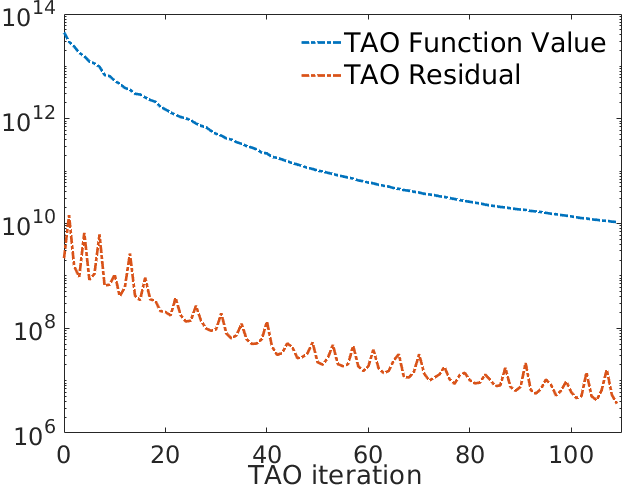}}
\caption{Inversion result in 2D}
\label{fig:real-inv}
\end{figure}

In Fig. \ref{fig:real-inv}, the diffusivity field after 50 and 110 TAO iterations are plotted as compared to the true diffusivity. The convergence history is also shown.

Due to the sparsity of the observations, the inversion could only reveals the diffusivity field at sample locations.

\begin{figure}[H]
\centering
\subfigure[Pressure at day 1]{\includegraphics[scale=0.25]{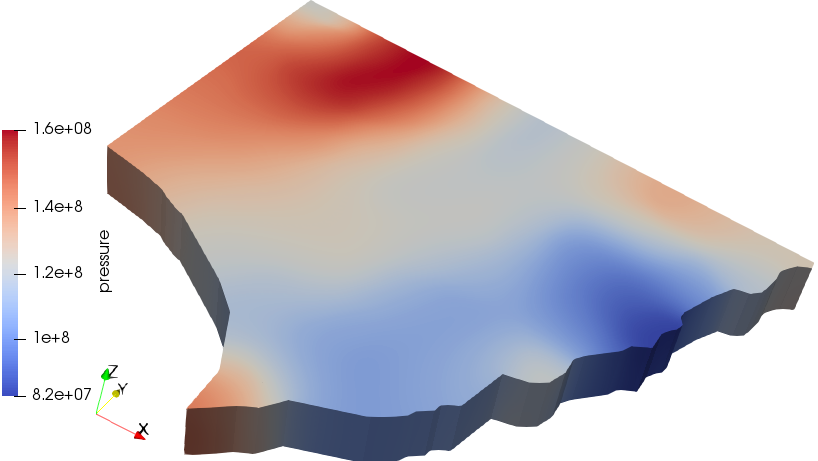}}
\subfigure[Pressure at day 200]{\includegraphics[scale=0.25]{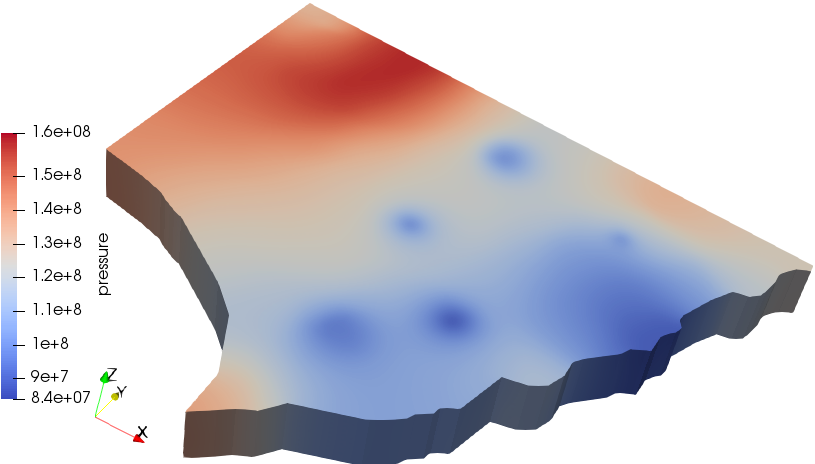}}
\subfigure[Mesh(tetrahedra)]{\includegraphics[scale=0.18]{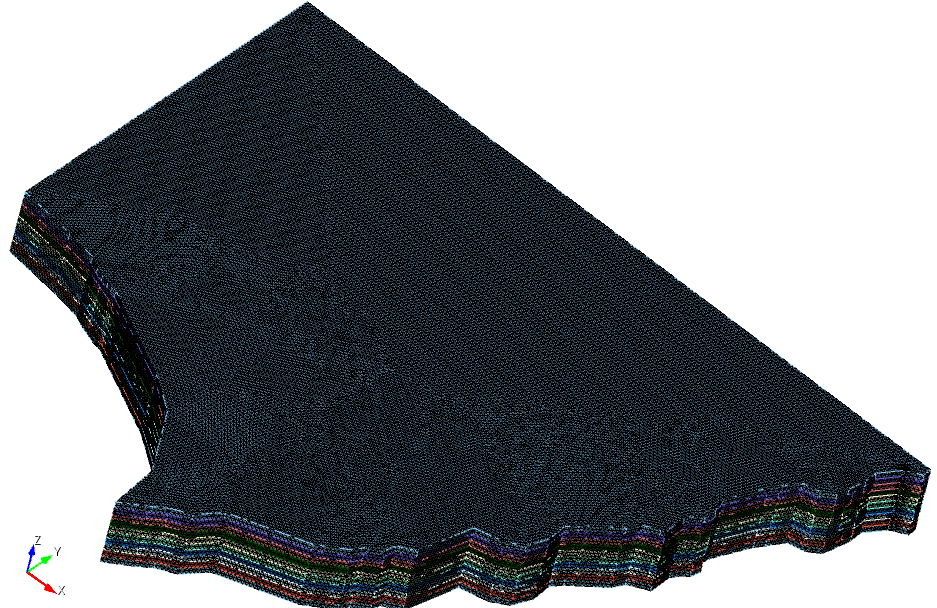}}
\label{fig:wtr3dicobs}
\caption{Pressure observed in 3D and corresponding mesh}
\end{figure}

3D example,
In the z direction, 2 kilometer. The initial condition is a steady state(run >1000 days from the funky ic I generated). Sinks are prescribed at 5 locations and run forwardly for 200 days. For the initial condition and observation data, please see Fig.\ref{fig:wtr3dicobs}

The parameters for this module is :
g=9.8m/$s^2$, $\rho=1.0\times 10^{3}$kg/$m^3$; $\phi=0.1$;$\beta=5.0\times 10^{-10}$ $Pa^{-1}$ and K=9.8$\times 10^{-11}$ m/s to 2.94$\times10^{-9}$m/s

\begin{figure}[H]
\centering
\subfigure[Iteration 40]{\includegraphics[scale=0.25]{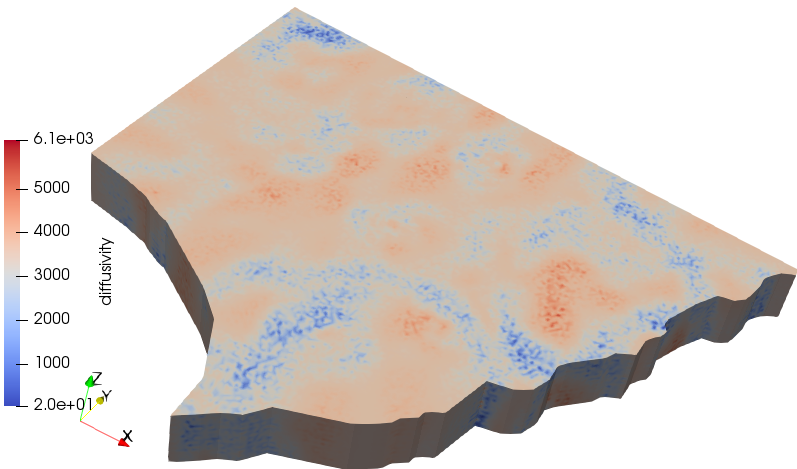}}
\subfigure[Iteration 100]{\includegraphics[scale=0.25]{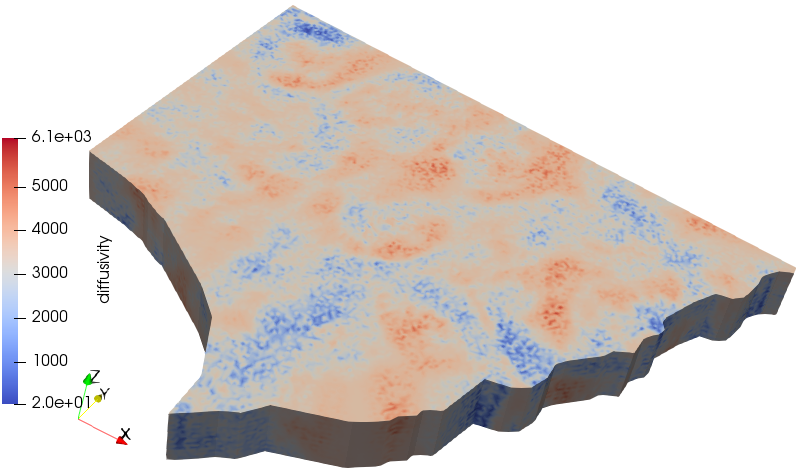}}\\
\subfigure[Iteration 160]{\includegraphics[scale=0.25]{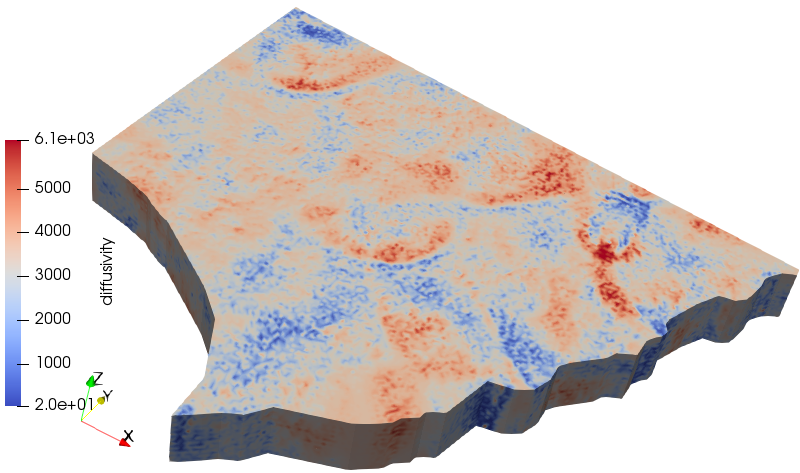}}
\subfigure[True diffusivity]{\includegraphics[scale=0.25]{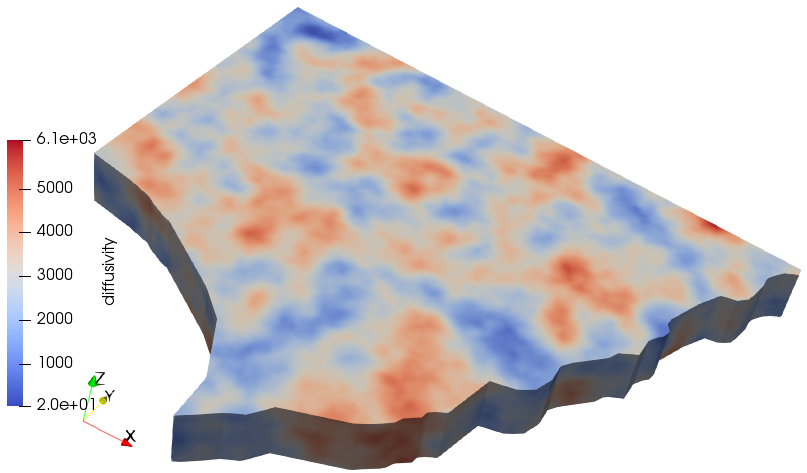}}
\caption{Inversion result in 3D}
\label{fig:wtr3dinv}
\end{figure}

Run the inversion on 64 nodes for 160 TAO iterations. The inverted diffusivity as compared to true distribution are shown in Fig. \ref{fig:wtr3dinv}.
And the convergence plot.
\begin{figure}[H]
\centering
{\includegraphics[scale=0.36]{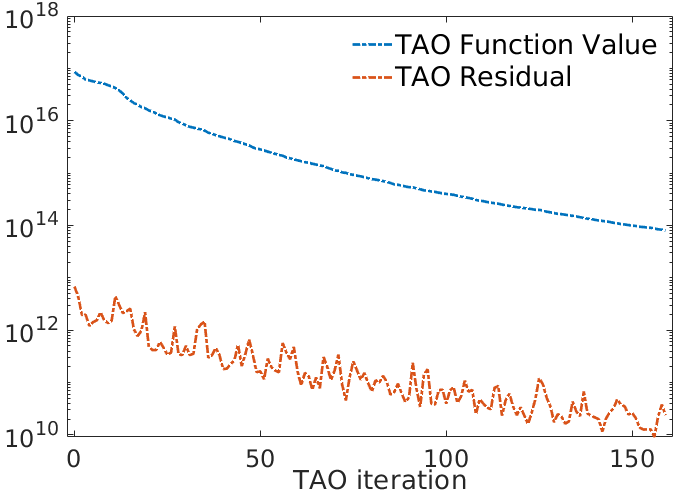}}
\caption{TAO convergence history}
\label{fig:wtr3dConv}
\end{figure}

\section*{Conclusions}

\section*{Acknowledgments}
SK thanks BER for support.
SW thanks LANL Parallel Computing Summer School for support. 

\newpage
\bibliographystyle{unsrt}
\bibliography{Shu_etal_2019}
\end{document}